\documentclass[cits]{PoS}
\usepackage[allowmove]{url}
\usepackage{breakurl}
\usepackage[square,numbers]{natbib}

\title{Multiwavelength observations of TANAMI sources}
\ShortTitle{Multiwavelength observations of TANAMI sources}
\author{\speaker{Felicia Krau{\ss}}\\
        Dr. Karl Remeis Observatory \& ECAP, Sternwartstr. 7, 96049
        Bamberg, Germany\\
        Julius-Maximilian Universit\"at W\"urzburg,
        Emil-Fischer-Str. 31, 97074 W\"urzburg, Germany\\
        E-mail: \email{Felicia.Krauss@sternwarte.uni-erlangen.de}}

\author{Cornelia M\"uller\\
         Dr. Karl Remeis Observatory \& ECAP, Sternwartstr. 7, 96049
        Bamberg, Germany\\
        Julius-Maximilian Universit\"at W\"urzburg,
        Emil-Fischer-Str. 31, 97074 W\"urzburg, Germany\\
        E-mail: \email{Cornelia.Mueller@sternwarte.uni-erlangen.de}
        }

\author{Matthias Kadler\\
        Julius-Maximilan Universit\"at W\"urzburg, 
        Emil-Fischer-Str. 31, 97074 W\"urzburg, Germany\\
        }

\author{J\"orn Wilms\\
        Dr. Karl Remeis Observatory \& ECAP, Sternwartstr. 7, 96049
        Bamberg, Germany\\
        }

\author{Moritz B\"ock\\
        Max-Planck-Institut f\"ur Radioastronomie,53010 Bonn, Germany\\
        }

\author{Roopesh Ojha\\
        NASA, GSFC, Greenbelt, MD 20771 USA\\
        }

\author{Eduardo Ros\\
        Universitat de Val\`{e}ncia, 46100 Burjasot, Spain\\
        Max-Planck-Institut f\"ur Radioastronomie, 53010 Bonn, Germany\\
        }

\author{on behalf of the TANAMI collaboration}
     
\usepackage{xspace}

\newcommand\XMMNewton{\textsl{XMM-Newton}\xspace}

\newcommand\Swift{\textsl{Swift}\xspace}
\newcommand\Fermi{\textsl{Fermi}\xspace}

\newcommand\ignoreit[1]{}

\abstract{The TANAMI VLBI program is monitoring a sample of 84 Active Galactic
Nuclei of the Southern Sky at 8.4 and 22 GHz.
The combination of VLBI and multiwavelength data allows us to study
changes in the spectral energy distributions, as well as changes in the
structure of the inner jets and to search correlations between both.\\
We present initial results of the multiwavelength analysis of a
sub-sample of the TANAMI sources, 
combining our radio data with simultaneous X-ray and optical/UV observations from \Swift and \XMMNewton, 
and gamma-ray data from \Fermi, focusing on the broadband spectral energy distributions as well as variability 
in different wavebands.}

\FullConference{11th European VLBI Network Symposium \& Users Meeting,\\
		October 9-12, 2012\\
		Bordeaux, France}

\begin{document}

\section{Introduction}

Active galactic nuclei (AGN) have been studied for over 100 years,
however, many questions about their evolution and emission mechanisms
remain to be answered.
VLBI measurements can resolve emission regions on milliarcsecond
scales which correspond to projected length scales of very few thousand Schwarzschild radii
from the central black hole for close AGN. Our goal is to study the
emission mechanisms of jets.
We combine VLBI radio data with
multiwavelength data from other instruments such as the \Swift
\textsl{Gamma-ray Burst mission} and the
\Fermi \textsl{Gamma-ray Space Telescope}. We use quasi-simultaneous data in order to
establish a catalog for spectral energy distributions of a large
sample of sources.

\section{The TANAMI Project}
\begin{figure}[!h]
    \centering
    \includegraphics[width=0.9\textwidth]{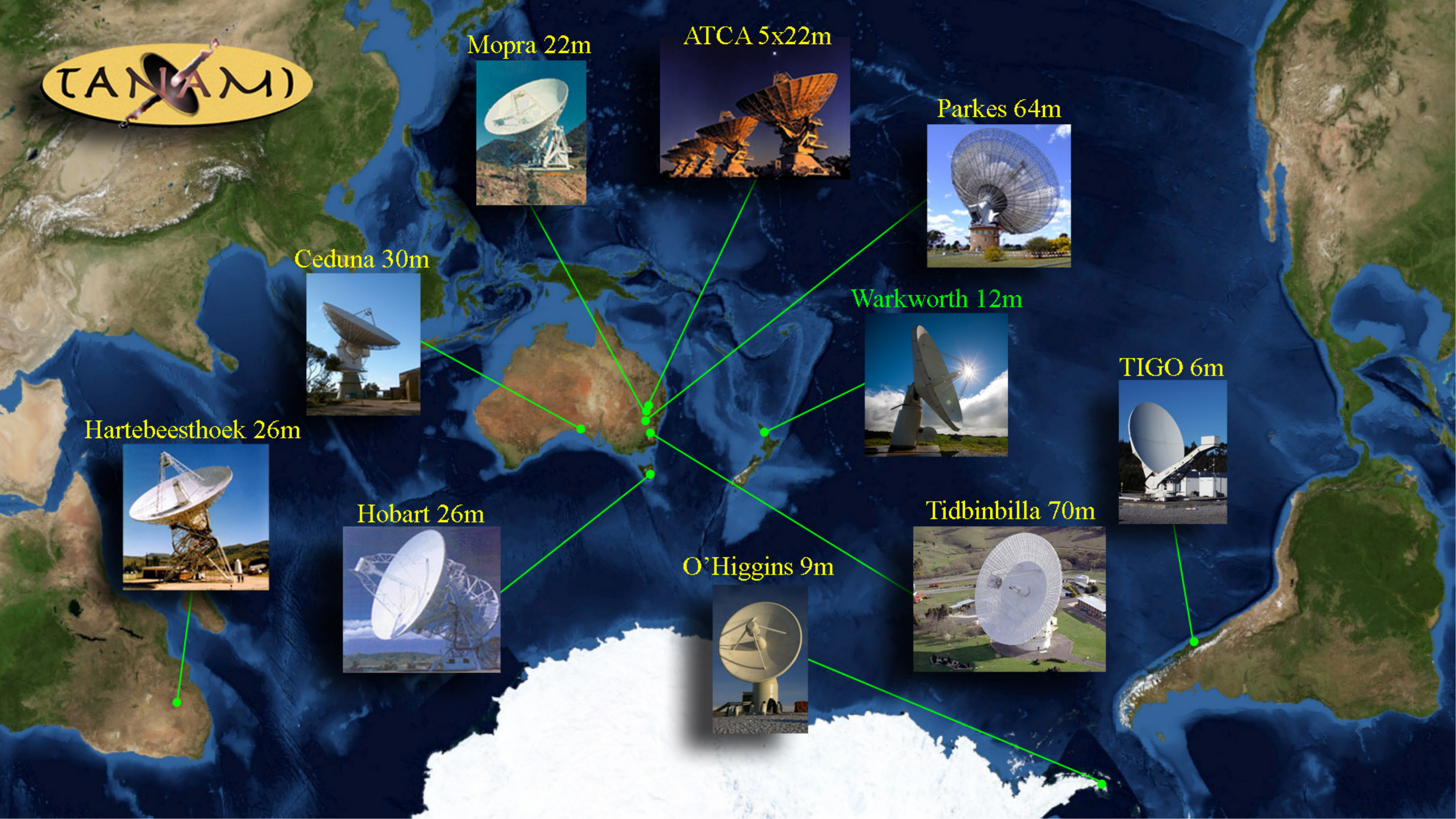}
    \caption{Map of the Southern Hemisphere showing the locations of the radio
      telescopes in the TANAMI Project.}
    \label{tanami01}
  \end{figure}

\textit{Tracking Active Galactic Nuclei with Austral Milliarcsecond
Interferometry} (TANAMI) is a VLBI program monitoring a sample of sources
south of -30 degrees declination since 2007, complementary to the
MOJAVE project \cite{2005AJ....130.1389L} on the
Northern Hemisphere. The initial sample of 43 sources has been expanded
to 84 since the launch of \Fermi in 2008 \citep{2011A&A...530L..11M}.
These sources are monitored every $\sim$3 months at 8.4 and 22 GHz
\citep{2010A&A...519A..45O}. Figure \ref{tanami01} shows a map of the
Southern Sky with the location of the TANAMI array.
TANAMI is using the Australian Long Baseline Array
(LBA+) with the antennas Parkes (64m), ATCA (5$\times$22\,m), Mopra (22\,m),
Hobart (26\,m), Ceduna (30\,m) and the associated antennas Tidbinbilla
(DSN 70\,m or 34\,m), the IVS antennas TIGO (6\,m, Chile) and O'Higgins (9\,m,
Antarctica), 
 Hartebeesthoek (26\,m, South Africa) and Warkworth (12\,m, New Zealand). %
For a detailed discussion of the TANAMI data and analysis see also \citep{muell13}.

\clearpage
\section{\Fermi/LAT light curve analysis and spectra}
AGN are very variable across the whole
electromagnetic spectrum. Information about the source state
is important to be able to create broadband spectra.
The \Fermi satellite is continuously monitoring the sky at $\gamma$-ray
energies since its launch in 2008. \Fermi observes in the 30 MeV to
300 GeV band \citep{2009ApJ...697.1071A}.
This allows us to obtain light
curves for all \Fermi/LAT detected sources (C. M\"uller et al., in
prep). 
55 of the TANAMI sources have been detected by \Fermi/LAT [4, 9].
As the \Fermi analysis is still ongoing we used the public light
curves which are available online. These light curves allow us to
identify high and low states of a
source.

\begin{figure}[h!]
  \centering
  \includegraphics[width=0.94\textwidth]{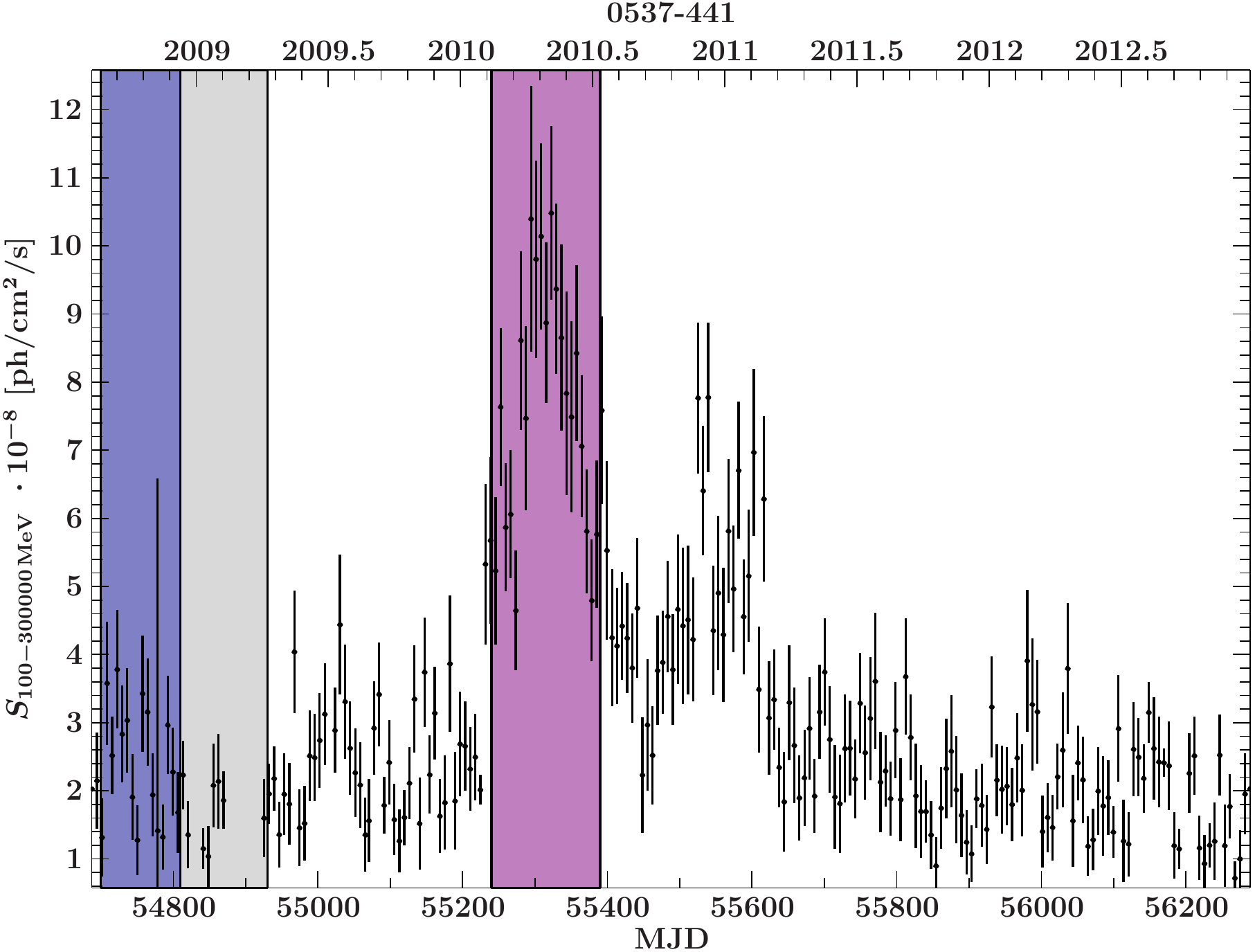}
  \caption{Identifying source states in the \Fermi light curve of PKS
    0537-441}
  \label{lc02}
\end{figure}

The $\gamma$-ray bright object PKS 0537-441 shows strong variability
in all wavebands. 
Figure \ref{lc02} shows its $\gamma$-ray light curve including a bright flare in 2010.

The colored areas show the time ranges used for
creating quasi-simultaneous broadband spectral energy distributions (SEDs).
The purple time range was identified as a high state,
the blue area as an intermediate state and the gray area as a low
state of the source.

The light curves show that quasi-simultaneous data are necessary for
broadband spectra. Here we used spectral $\gamma$-ray information from the
LAT 2-year point source catalog which contains flux information
summed over the first 2 years of \Fermi.
The energy range of these spectra are in the 30\,MeV -- 300\,GeV Band.
In the future we plan to create quasi-simultaneous spectra using
\Fermi data only in the time ranges given by the source state.

\section{Multiwavelength Spectral Energy Distributions}

The broadband spectral energy distribution (SED) for PKS 0537-441 has been created from
quasi-simultaneous TANAMI data (Kadler et al., in prep.) as well as
\Swift data (Optical/UV and X-ray) [5, 6]. 
The optical data have not been corrected for reddening and the \Fermi/LAT
data have not been corrected for the Extragalactic Background
Light (EBL) absorption.

In order not to falsify the data, the optical and X-ray data are not unfolded but modeled in detector space.
The modeling of the SEDs with log parabolas is still ongoing and
will be presented in a TANAMI SED catalog (Krau{\ss} et al., in prep.). 

    \begin{figure}[!ht]
      \centering
      \includegraphics[width=0.9\textwidth]{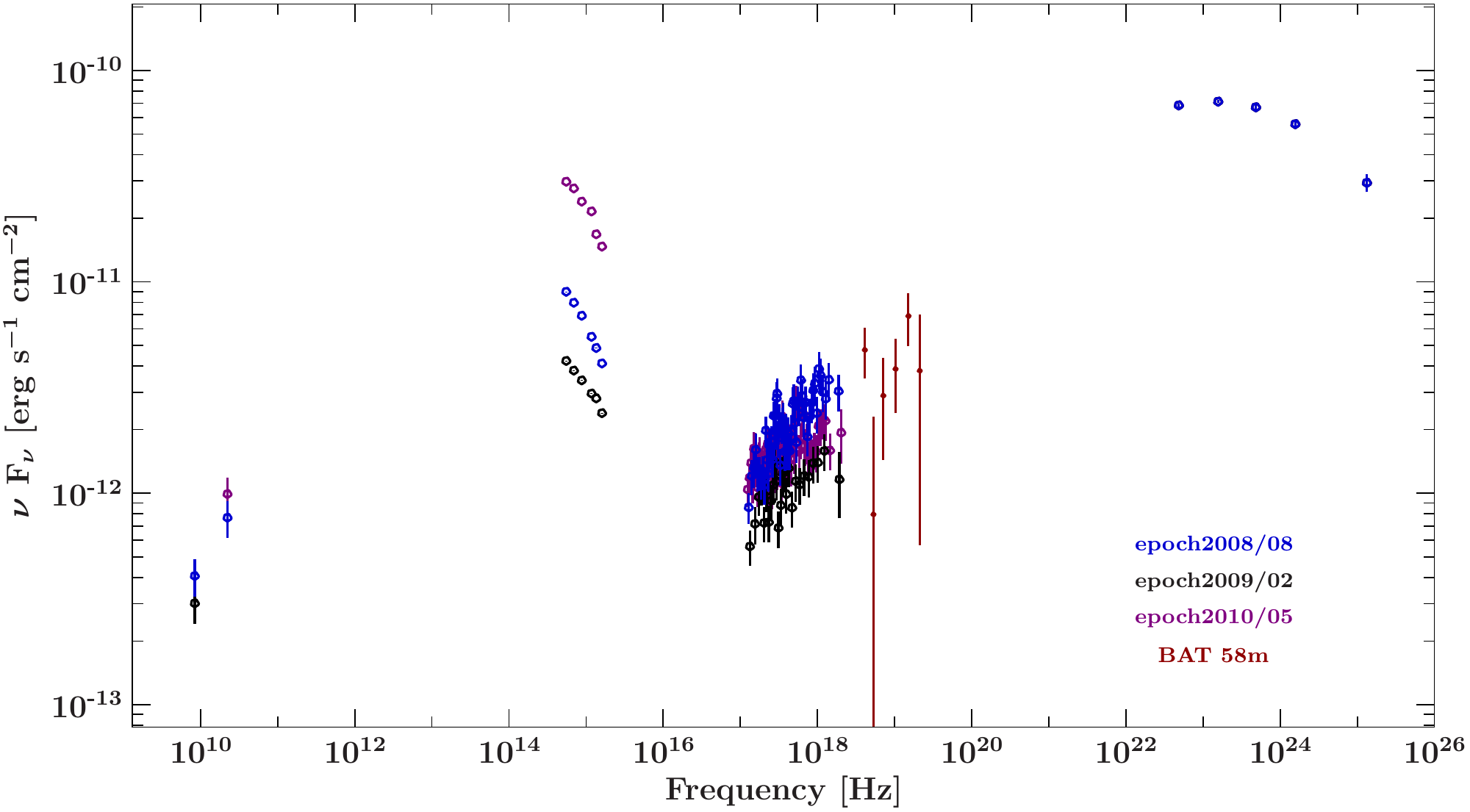}%
    \caption{Spectral Energy Distribution of PKS 0537-441 including
      non-simultaneous 58-month \Swift/BAT and 2FGL \Fermi/LAT data}
    \label{sed01}
    \end{figure}

The colors of the TANAMI, \Swift/UVOT and \Swift/XRT data in Figure
\ref{sed01} correspond to the
colors used for identifying the source states in Figure \ref{lc02}.
This figure also includes \Swift/BAT data from the 58 month
catalog [2, 3] and the \Fermi data
from the publicly available 2FGL catalog.
\cite{2012ApJS..199...31N}.
The \Swift/BAT and the \Fermi data are not simultaneous as both data sets
are summed over a large time range.
Figure \ref{sed01} also shows the variability that has been seen in
the light curve. This is very apparent in the optical regime.
The SED shows the double humped spectral form that is
usually seen in blazars.

The low-energy peak is generally interpreted as synchrotron
emission of electrons being accelerated in magnetic fields.
The high-energy peak is usually
associated with Synchrotron Self-Compton (SSC) emission. An
alternative are hadronic or lepto-hadronic models.
Spectral properties can be connected to the TANAMI VLBI information
with two analysis methods.
High-resolution images at two frequencies can be used to created
spectral index maps of jets to pin down emission regions which are
possible emission sites for $\gamma$-rays \citep{muell13}.
The continuous monitoring of the sources at radio and $\gamma$-ray
frequencies allows us to study the radio-$\gamma$-ray connection even
further. 
The ejection of jet features, revealed as traveling features at
milliarcsecond resolution can be correlated with high states seen by \Fermi
and possible frequency-dependent time lags can be discerned \citep{2004ApJ...613..725S}.

\section{Results and future work}

We have given a short overview of the TANAMI project and presented initial
results of a multiwavelength study of southern jets.
This includes quasi-simultaneous broadband spectral energy
distributions using TANAMI, \Swift/UVOT and \Swift/XRT data.
Non-simultaneous \Swift/BAT and \Fermi data have been added to the
spectral energy distributions.
Analysis of the publicly available \Fermi light curves
show the need
for quasi-simultaneous coverage in all wavebands in order to study the
different source states.
Blazar variability and spectral energy distributions have to be
studied in detail in the future and applied to a large source sample.

For a better understanding of the emission mechanisms of jets in active
galactic nuclei more simultaneous multiwavelength campaigns are
needed.

\section{Acknowledgments}

We thank the Deutsche Forschungsgemeinschaft for support under grant
WI 1860/10-1.
This research has made use of data and/or software provided by the
High Energy Astrophysics Science Archive Research Center (HEASARC),
which is a service of the Astrophysics Science Division at NASA/GSFC
and the High Energy Astrophysics Division of the Smithsonian
Astrophysical Observatory.
This research has made use of the NASA/IPAC Extragalactic Database
(NED) which is operated by the Jet Propulsion Laboratory, California
Institute of Technology, under contract with the National Aeronautics
and Space Administration.
This research has made use of the SIMBAD database,
operated at CDS, Strasbourg, France.

\end{document}